\documentclass[12pt]{article}
\usepackage{graphics} 
\usepackage{cite}
\usepackage{epsfig}
\usepackage{color}

\newcommand{\beq}{\begin{equation}}
\newcommand{\eeq}{\end{equation}}
\newcommand{\beqa}{\begin{eqnarray}}
\newcommand{\eeqa}{\end{eqnarray}}
\newcommand{\bea}{\begin{eqnarray}}
\newcommand{\eea}{\end{eqnarray}}

\newcommand{\canal}[2] {$\!(#1\!\rightarrow\!#2)$}
\newcommand{\noncanal} {$\!(\textrm{non-canalyzing})$}
\newcommand{\AND}  {\textrm{ \sc and  }}
\newcommand{\OR}   {\textrm{ \sc or   }}
\newcommand{\XOR}  {\textrm{ \sc xor  }}
\newcommand{\NOT}  {\textrm{\sc not  }}

\newcommand{\TRUE} {\textrm{\sc true}}
\newcommand{\FALSE}{\textrm{\sc false}}
\newcommand{\nobs} {n_{\textrm{\scriptsize obs}}}
\newcommand{\ncalc}{n_{\textrm{\scriptsize calc}}}
\newcommand{\Odef} {O_{\textrm{\scriptsize default}}}
\newcommand{\implies}{\Rightarrow}
\newcommand{\Pth}{P_\textrm{\scriptsize th}}

\begin{document}

\begin{flushright}
LU TP 03-21\\
June 29, 2003
\end{flushright}

\vspace{0.2in}

\begin{center}

\Large
{\bf Random Boolean Network Models}\\
{\bf and the Yeast Transcriptional Network}

\vspace{0.2in}
\large

Stuart Kauffman$^1$, Carsten Peterson$^2$,\\ 
Bj\"orn Samuelsson$^2$ and Carl Troein$^2$

\vspace{0.25in}

\normalsize 

$^1$Department of Cell Biology and Physiology, \\ 
University of New Mexico Health Sciences Center, Albuquerque, NM 87131, USA

\vspace{0.15in}
$^2$Complex Systems Division, Department of Theoretical Physics\\
Lund University,  S\"{o}lvegatan 14A,  S-223 62 Lund, Sweden \\
{\tt http://www.thep.lu.se/complex/}

\vspace{0.3in}

Submitted to {\it Proceedings of the National Academy of Sciences USA}

\end{center}

\vspace{0.60in}

\normalsize

\noindent
{\it Corresponding author:} C. Peterson; carsten@thep.lu.se
%\makebox[43mm]{}tel: 505 984-1998 \\ 
%\makebox[43mm]{}fax: 505 984-8245 \\
%\makebox[43mm]{}e-mail: SKauffman@salud.unm.edu

\vspace{0.1in}
\noindent
{\it Subject Category:} Biological Sciences: Biophysics.

\vspace{0.1in}
\noindent
{\it Keywords}: genetic networks, yeast, Boolean rules, dynamical systems.

\vspace{0.1in}
\noindent
{\it Manuscript information:} 6 text pages, 4 figures and 0 tables; 
95 words in abstract; 29248 total characters in the paper; supporting 
information for the web.
%22098 chars
%figures: 180 * __ cm = 180 * (6+9+6+6+6) = 5940
%fig.extra: 120 * 5 figs = 600
%equations: 60 * __ lines = 60 * 7 = 420
%eq.extra: 120 * 2 eqs = 240
%total: 22098 + 5940 + 600 + 420 + 240 = 29298

\newpage

\begin{center}
{\large \bf Abstract}
\end{center}

\noindent

The recently measured yeast transcriptional network is analyzed in
terms of simplified Boolean network models, with the aim of determining 
feasible rule structures, given the requirement of stable 
solutions of the generated Boolean networks. We find that for 
ensembles of generated models, those with canalyzing Boolean rules 
are remarkably stable, whereas those with random Boolean 
rules are only marginally stable. Furthermore, substantial parts of the 
generated networks are frozen, in the sense that they reach the same
state regardless of initial state. Thus, our ensemble approach 
suggests that the yeast network shows highly ordered dynamics.

\newpage

\section{Introduction}

The regulatory network for {\it Saccharomyces cerevisiae} was recently 
measured  \cite{Young:02} for 106 of the 141 known transcription factors by 
determining the bindings of transcription factor proteins to promoter 
regions on the DNA. Associating the promoter regions with genes yields 
a network of directed gene-gene interactions.
As described in \cite{Young:02,Hughes:02} the significance of measured 
bindings with regard to infering putative interactions are quantified 
in terms of $P$ values. The authors of \cite{Young:02} 
did not infer interactions having $P$ values above a threshold value 
$\Pth=0.001$ for most of their analysis. Small threshold
values $\Pth$ correspond to a small number of inferred interactions
with high quality, whereas larger values correspond to more inferred
connections but of lower quality. It was found that for the
$\Pth=0.001$ network, the fan-out from each transcription factor to 
its regulated targets is substantial, on the average 38 \cite{Young:02}.
From the underlying data (website: http://web.wi.mit.edu/young/regulatory\_network) 
one finds that fairly few signals feed into each of them; on the average 1.9.
The experiments yield the regulatory network architecture but neither the
interaction rules at the nodes, nor the dynamics of the system, nor 
its final states.

With no direct experimental results on the states of the system, 
there is of course
no systematic method to pin down the interaction rules, not even within
the framework of simplified and coarse-grained genetic network models, e.g.
ones where the rules are Boolean. One can nevertheless attempt to investigate
to what extent the measured architecture can select between classes
of Boolean models \cite{Kauffman:93}, based upon criteria of stability. 

We generate ensembles of different model networks on the given
architecture, and analyze their behavior with respect to
stability. In a stable system small initial perturbations should not 
grow in time. This is investigated by monitoring how the 
Hamming distances between different initial states evolve in a 
``Derrida plot'' \cite{Derrida:86}. If small Hamming distances
diverge in time, the system is unstable and vice versa. Based upon
this criterion we find that synchronously updated random Boolean
networks (with a flat rule distribution) are marginally stable on
the transcriptional network of yeast.

Using a subset of Boolean rules, {\it nested canalyzing functions}
(see sect.~\ref{seq:genrul}), the ensemble of networks exhibits
remarkable stability. The notion of nested canalyzing functions is
introduced to provide a natural way of generating canalyzing
rules, which are abundant in biology
\cite{Harris:02}. Furthermore, it turns out that for these networks,
there exists a fair amount of {\it forcing structures} \cite{Kauffman:93},
where non-negligible parts of the networks are frozen to fixed final states
regardless of the initial conditions. Also, we investigate the
consequences of rewiring the network while retaining the local
properties; the number of inputs and outputs for each node
\cite{Maslov:02}.

To accomplish the above, some novel tools and techniques were developed 
and used. In order to include more interactions than those in the
$\Pth=0.001$ network \cite{Young:02}, we investigate how network
properties, local and global, change as $\Pth$ is increased. We find a
transition slightly above $\Pth=0.005$, indicating the onset of noise
in the form of biologically irrelevant inferred connections.
In \cite{Harris:02} extensive literature studies revealed that,
for eukaryotes, the rules seem to be canalyzing. We develop a
convenient method to generate a distribution of canalyzing
rules, that fits well with the list of rules in \cite{Harris:02}.

\section{Methods and Models}

\subsection{Choosing Network Architecture}

In \cite{Young:02}, $P$ values were calculated as measures of
confidence in the presence of an interaction. With further
elucidation of noise levels, one might increase the threshold for $P$
values from the value 0.001 used in \cite{Young:02}. To this end we
compute various network properties, to investigate if there is any
value of $\Pth$ for which these properties exhibit a transition that
can be interpreted as the onset of noise. In Fig.~1 the number of
nodes, mean connectivity, mean pairwise distance (radius) and fraction
of node pairs connected are shown. As can be seen, there appears to be
a transition slightly above $\Pth=0.005$. In what follows we therefore
focus on the network defined by $\Pth=0.005$. Furthermore, we (recursively)
remove genes which have no outputs to other genes, since these are not
relevant for the network dynamics. The resulting network is shown in
Fig.~2.
 
\subsection{Generating Rules}
\label{seq:genrul}

In \cite{Young:02}, the architecture of the network is determined, but
not the specific rules for the interactions. In order to investigate
the dynamics on the measured architecture, we repeatedly assign
a random Boolean rule to each node in the network. We use two rule
distributions; one null-hypothesis and one distribution that
agrees with rules compiled from the literature \cite{Harris:02}
(see {\it Supporting Information}). In both cases we ensure that
every rule depends on all of its inputs, since the dependence should
be consistent with the network architecture.

As a null-hypothesis, we use a flat distribution among all Boolean
functions that depend on all inputs. For rules with a few inputs, this
will create rules that can be expressed with normal Boolean functions
in a convenient way. In the case of many inputs, most rules are
unstructured and the result of toggling one input value will
appear random.

In biological systems, the distribution of rules is likely to be 
structured. Indeed all of the compiled rules in \cite{Harris:02} 
are canalyzing \cite{Kauffman:93}; a canalyzing Boolean function
\cite{Kauffman:93} has at least one input, such that for at least one
input value, the output value is fixed.
It is not straightforward to generate biologically relevant canalyzing
functions. 
A canalyzing rule implies some structure, but the function of the 
non-canalyzing inputs (when the canalyzing inputs are clamped to their 
non-canalyzing values) could be as disordered as the full set of random 
Boolean rules. However, the canalyzing structure is repeated in a nested
fashion for almost all rules in \cite{Harris:02}. Hence,
we introduce the concept of {\it nested canalyzing functions}
(see {\it Appendix}), which can be used to generate 
distributions of canalyzing rules. Actually, of the 139 rules in
\cite{Harris:02} only 6 are not 
nested canalyzing functions (see {\it Supporting Information}).

A special case of nested canalyzing functions is the recently
introduced notion of \textit{chain functions} \cite{GatViks:03}
(see {\it Appendix}). Chain functions are the most abundant form
of nested canalyzing functions, but 32 of the 139 rules in
\cite{Harris:02} fall outside this class.

It turns out that the rule distribution of nested canalyzing
functions in \cite{Harris:02}
can be well described by a model with only one parameter
(see {\it Appendix}). Hence, we use this model to mimic the compiled rule
distribution. The free parameter determines the degree of asymmetry
between active and inactive states, and its value reflects the fact that
most genes are inactive at any given time in a gene regulatory system.

\subsection{Analyzing the Dynamics}

A biological system is subject to a substantial amount of noise,
making robustness a necessary feature of any model. We expect a
transcriptional network to be stable, in that a random disturbance can
not be allowed to grow uncontrollably. Gene expression levels can
be approximated as Boolean, as genes tend to be either active or
inactive.  This approximation for genetic networks is presumably
easier to handle for stability issues than for general dynamical
properties. Using synchronous updates is computationally and
conceptually convenient though it may at first sight appear
unrealistic. However, in instances of strong stability, the update
order should not be very important.

To study the time development of small fluctuations in this discrete
model with synchronous updating, we investigate how the Hamming
distance between two states evolves with time.  In a Derrida plot
\cite{Derrida:86} pairs of initial states are sampled at defined
initial distances $H(0)$ from the entire state space, and their
mean Hamming distance $H(t)$ after a fixed time $t$ is plotted
against the initial distance $H(0)$. The slope in the low-$H$ region
indicates the fate of a small disturbance.  If the curve is
above/below the line $H(t) = H(0)$ it reflects instability/stability
in the sense that a small disturbance tend to increase/decrease during
the next $t$ time steps (see Fig.~3).

It is not uncommon that transcription factors control their own
expression. In some cases genes up-regulate themselves, with the
effect that their behavior becomes less linear and more
switch-like. This is readily mimicked in a Boolean network. However,
in the other case, where a transcription factor down-regulates
itself, the system will be stabilized in a model with continuous
variables, provided that the time delay of the self-interaction is not
too large. Boolean networks can only model the limit of large time
delays, which gives rise to nodes that in an unbiological manner repeatedly
flip between no activity and full activity without requiring any
external input. Thus, the self-interactions need to be treated
as a special case in the Boolean approximation. To this end, we
consider three different alternatives:
\begin{enumerate}
\item View the self-interactions as internal parts of the rules;
all self-interactions are removed.
\label{self removed}
\item Remove the possibility for self-interactions to be down-regulating. 
\label{self switch}
\item No special treatment of self-interactions.
\label{self all}
\end{enumerate}
It is natural to use alternative \ref{self removed} as a reference point
in order to understand the effect of the self-interactions in alternatives
\ref{self switch} and \ref{self all}.

We want to examine how the geometry of networks influence the
dynamics. It is known \cite{Kauffman:93} that the distributions of in-
and out-connectivities of the nodes strongly affect the dynamics in
Boolean networks, but how important is the overall architecture? If
for each node we preserve the connectivities, but otherwise rewire the
network randomly \cite{Maslov:02}, how is the dynamics affected?  
For a Derrida plot with $t = 1$, there is no change. If we only take a 
single time step from a random state, the outputs will not have time to be
used as inputs. There will be correlations between nodes, but the measured
quantity $H(1)$ is a mean over all nodes, and this is not affected by
these correlations. Hence, $H(1)$ is not changed by the rewiring. In
order to get a better picture of the dynamics we need to increase
$t$. However, if we go high enough in $t$ to probe larger structures
in the networks, we lose sight of the transient effects of a
perturbation.

To remedy this, we opt to select a fixed initial Hamming distance
$H(0)$, and examine the expectation value of the distance as a
function of time, using the nested canalyzing rules. As noise
entering the biological network would act on the current state of the
system rather than on an entirely random one, we select one of the
states to be a fixed point of the dynamics, and let the probability of
any given fixed point be proportional to the size of its attractor
basin. A graph of $H(t)$ shows the relaxation behavior of the
perturbed system where the self-interactions have been removed (see
Fig.~4a). We investigate the role of the self-interactions
both in terms of relaxation of a perturbed fixed point (see
Fig.~4b) and in terms of probabilities for random trajectories
to end up in distinct fixed points and cycles.

The assumption that the typical state of these networks is a fixed
point can be motivated. A forcing connection \cite{Kauffman:93} is a pair of
connected nodes, such that control over a single input to one node
is sufficient to force the output of the other node to one of the
Boolean values. With canalyzing rules, this is fulfilled when the
canalyzed output of the first node is a canalyzing input to the
second. The existence of forcing structures implies stability, as
a (forcing) signal traveling through such a structure will block
out other inputs and is thereby likely to cause information loss.
Abundant forcing structures should tend to favor fixed points.

\section{Results and Discussion}

Despite absence of knowledge about initial and final states, we have
been able to get a hint about possible interaction rules within a
Boolean network framework for the yeast transcriptional network. Our 
findings are: 

\begin{itemize}

\item Canalyzing Boolean rules confer far more stability than rules drawn
from a flat distribution as is clear from the Derrida plots in Fig.~3. 
Yet, even a flat distribution of Boolean functions yields
marginal stability.

\item The dynamical behavior around fixed points is more stable for 
the measured network than for the rewired ones, though only in the early 
time evolution (2--3 time steps) of the systems (see Fig.~4a). The behavior at 
this time scale can be expected to depend largely on small network motifs, 
whose numbers are systematically changed by the rewiring \cite{Maslov:02}. 

\item The removal of self-couplings increases the stability in
these networks. However, the relaxation is only changed significantly
if we allow the toggling of self-interacting nodes (see Fig.~4b). This
means that a node with a switch-like self-interaction is not likely to
be toggled by its inputs during the relaxation. Nor do the
down-regulating self-interactions alter the relaxation. This means
that the overall properties of relaxation to fixed points can be
investigated regardless of how the self-interactions should be
modeled.

\item The number of attractors and their length distribution are strongly
dependent on how the self-interactions are modeled. The average numbers
of distinct fixed points per rule assignments found in 1000 trials of
different trajectories are 1.02, 4.33 and 3.79, respectively, for the three
self-interaction models. The numbers of 2-cycles are 0.02, 0.09 and
0.38, respectively. Longer cycles are less common; in total they
sum up to 0.03, 0.11 and 0.11, respectively.

\item Forcing structures \cite{Kauffman:93} are prevalent for this architecture 
with canalyzing rules, as is evident from Fig.~2. On average 56\,\% of the 
couplings belong to forcing structures. As a consequence, most nodes will be
forced to a fixed state regardless of the initial state of the
network. Even the highly connected nodes (in the center of the network)
will be forced to a fixed state for a vast majority of the random rule
assignments. In most cases, the whole network will be forced to a
specific fixed state. At first glance this might seem un-biological.
However, in the real world there are more inputs to the system than the
measured transcription factors, and to study a process such as the cell
cycle, one may need to consider additional components of the system.
With more inputs such a strong stability --- of the measured part of
the network --- may be necessary for robustness of the entire system.

\end{itemize}

Future reverse engineering projects in transcriptional networks 
may be based on the restricted pool of nested canalyzing rules,
which have been shown to generate very robust networks in this case.
It should be pointed out that the notion of nested canalyzing functions
is not intrinsically Boolean. For instance, the same concept can
be applied to nested sigmoids.

\subsection*{Acknowledgments}

We thank Stephen Harris for providing us with details underlying reference 
\cite{Harris:02}.
This research was initiated at the Kavli Institute for Theoretical Physics 
in Santa Barbara (CP and SK) and was supported in part by the National Science
Foundation under Grant No.~PHY99-07949. CT acknowledges the support from the 
Swedish National Research School in Genomics and Bioinformatics.

\newpage

\subsection*{Appendix: Nested Canalyzing Functions}

The notion of nested canalyzing functions is a natural extension of
canalyzing functions. Consider a $K$-input Boolean rule $R$ with
inputs $i_1, \ldots, i_K$ and output $o$. $R$ is canalyzing on the
input $i_m$ if there are Boolean values $I_m$ and $O_m$ such that $i_m
= I_m \implies o = O_m$. $I_m$ is the canalyzing value, and
$O_m$ is the canalyzed value for the output.

For each canalyzing rule $R$, renumber the inputs in a way such that
$R$ is canalyzing on $i_1$. Then, there are Boolean values $I_1$ and
$O_1$ such that $i_1 = I_1 \implies o = O_1$. To investigate the case
$i_1 = \NOT I_1$, fix $i_1$ to this value. This defines a new rule
$R_1$ with $K-1$ inputs;
$i_2, \ldots, i_K$. In most cases, when picking $R$ from compiled data,
$R_1$ is also canalyzing. Then, renumber the inputs in order for $R_1$
to be canalyzing on $i_2$. Fixing $i_2 = \NOT I_2$ renders a rule
$R_2$ with the inputs $i_3, \ldots, i_K$. As long as the rules $R,
R_1, R_2, \ldots$ are canalyzing, we can repeat this procedure until
we find $R_{K-1}$ which has only one input $i_K$ and hence is
trivially canalyzing. Such a rule $R$ is a {\it nested canalyzing
function} and can be described by the canalyzing input values $I_1,
\ldots, I_K$ together with their respective canalyzed output values
$O_1, \ldots, O_K$ and an additional value $\Odef$. The output is
given by
\[
  o = \left\{\begin{array}{ll}
    O_1 & \textrm{if } i_1 = I_1 \\
    O_2 & \textrm{if } i_1 \neq I_1 \AND i_2 = I_2 \\
    O_3 & \textrm{if } i_1 \neq I_1 \AND i_2 \neq I_2 \AND i_3 = I_3 \\
     \vdots   & \\
    O_K & \textrm{if } i_1 \neq I_1 \AND \cdots \AND i_{K-1} \neq I_{K-1}
           \AND i_K = I_K \\
    \Odef & \textrm{if } i_1 \neq I_1 \AND \cdots \AND i_K \neq I_K~. \\
  \end{array} \right.
\]
The notion of {\it chain functions} in \cite{GatViks:03} is equivalent to
nested canalyzing functions that can be written on the form
$I_1 = \cdots = I_{K-1} = \FALSE$.

We want to generate a distribution of rules with $K$ inputs, such that
all rules depend on every input. The dependency requirement is
fulfilled if and only if $\Odef = \NOT O_K$. Then, it remains to
choose values for $I_1, \ldots, I_K$ and $O_1, \ldots, O_K$. 
These values are independently and randomly chosen with the probabilities
\[
  p(I_m = \TRUE) = p(O_m = \TRUE) = \frac{\exp(-2^{-m}\alpha)}
         {1 + \exp(-2^{-m}\alpha)}
\]
for $m = 1,\ldots,K$. For all generated distributions, we let $\alpha
= 7$.

The described scheme is sufficient to generate a well-defined rule
distribution, but each rule has more than one representation in $I_1,
\ldots, I_K$ and $O_1, \ldots, O_K$. In {\it Supporting
Information} we describe how to obtain a unique representation, which
is applied to the rules compiled in \cite{Harris:02}. This
enables us to present a firm comparison between the generated distribution
and the list of rules in \cite{Harris:02}.

{}

\newpage

\parindent 0pt
\parskip 3ex

{\Large\bf Figure captions}

%{\bf Fig.~1.}\quad 
%Topological properties of the yeast regulatory network in \cite{Young:02} 
%for different $P$ value thresholds: Number of nodes (solid black), 
%mean connectivity (dotted blue), mean pairwise distance [radius]
%(dotted-solid red) and fraction of node pairs that are connected
%(dashed green). The right $y$-axis corresponds to the number
%of nodes with no outputs, whereas the other quantities are indicated on
%the left $y$-axis. Self-couplings were excluded, but the figure looks
%similar when they are included. The dashed black line marks the
%threshold $\Pth=0.005$.

{\bf Fig.~1.}\quad 
Topological properties of the yeast regulatory network in \cite{Young:02} 
for different $P$ value thresholds: Number of nodes (solid line), 
mean connectivity (dotted line), mean pairwise distance [radius]
(dotted-solid line) and fraction of node pairs that are connected
(dashed line). The right $y$-axis corresponds to the number
of nodes with no outputs, whereas the other quantities are indicated on
the left $y$-axis. Self-couplings were excluded, but the figure looks
similar when they are included. The dashed vertical line marks the
threshold $\Pth=0.005$.

%{\bf Fig.~2.}\quad
%The $\Pth=0.005$ network excluding nodes with no outputs to other
%nodes than itself. The filled areas in the arrow-heads are
%proportional to the probability of each coupling to be in a forcing
%structure when the nested canalyzing rules are used on the
%network without self-interactions. This probability ranges from
%approximately 1/4 for the inputs to YAP6 to 1 for the inputs to
%one-input nodes. Nodes that will reach a frozen state (on or off)
%in the absence of down-regulating self-interactions, regardless
%of the choice of rules, are shown in dashed green. For the other
%nodes, the color indicates the probability of being frozen in the
%absence of self-interactions, ranging from just under 97\,\% (black)
%to over 99.9\,\% (light blue).

{\bf Fig.~2.}\quad
The $\Pth=0.005$ network excluding nodes with no outputs to other
nodes than itself. The filled areas in the arrow-heads are
proportional to the probability of each coupling to be in a forcing
structure when the nested canalyzing rules are used on the
network without self-interactions. This probability ranges from
approximately 1/4 for the inputs to YAP6 to 1 for the inputs to
one-input nodes. Nodes that will reach a frozen state (on or off)
in the absence of down-regulating self-interactions, regardless
of the choice of rules, are shown in dashed. For the other
nodes, the grey scale indicates the probability of being frozen in the
absence of self-interactions, ranging from just under 97\,\% (bold black)
to over 99.9\,\% (gray).

%{\bf Fig.~3.}\quad 
%Evolution of different Hamming distances $H(0)$ with one time step to
%$H(1)$ (Derrida plots \cite{Derrida:86}) for random rules (pink) and
%nested canalyzing rules (blue) with and without self-couplings (dashed
%borders) respectively. (Down-regulating self-couplings are 
%allowed.) The bands correspond to 1$\sigma$ variation
%among the different rule assignments generated on the architecture in
%Fig.~2. Statistics were gathered from 1000 starts on each of 1000
%rule assignments.

{\bf Fig.~3.}\quad 
Evolution of different Hamming distances $H(0)$ with one time step to
$H(1)$ (Derrida plots \cite{Derrida:86}) for random rules (dark grey) and
nested canalyzing rules (light grey) with and without self-couplings (dashed
borders) respectively. (Down-regulating self-couplings are 
allowed.) The bands correspond to 1$\sigma$ variation
among the different rule assignments generated on the architecture in
Fig.~2. Statistics were gathered from 1000 starts on each of 1000
rule assignments.

%{\bf Fig.~4.}\quad
%The average time evolution of perturbed fixed points for nested
%canalyzing rules, starting from Hamming
%distance $H(0)=5$; (a) impact of the network architecture
%and (b) impact of the self-interactions. The blue lines
%(marked with circles) in both figures correspond to the network in
%Fig.~2 without self-interactions. The red lines in (a) show the
%relaxation for 26 different rewired architectures with no
%self-interactions, with 1$\sigma$ errors of the calculated
%means indicated by the line widths. The black lines in (b) correspond to the
%network in Fig.~2 with self-interactions. The upper line shows the case
%when it is allowed to toggle nodes with self-interactions as a state
%at $H(0)=5$ is picked, while the lower line shows the relaxation if
%this is not allowed. The widths of these lines show the difference
%between allowing self-interactions to be repressive or not.

{\bf Fig.~4.}\quad
The average time evolution of perturbed fixed points for nested
canalyzing rules, starting from Hamming
distance $H(0)=5$; (a) impact of the network architecture
and (b) impact of the self-interactions. The lines marked  
with circles in both figures correspond to the network in
Fig.~2 without self-interactions. The grey lines in (a) show the
relaxation for 26 different rewired architectures with no
self-interactions, with 1$\sigma$ errors of the calculated
means indicated by the line widths. The black lines in (b) correspond to the
network in Fig.~2 with self-interactions. The upper line shows the case
when it is allowed to toggle nodes with self-interactions as a state
at $H(0)=5$ is picked, while the lower line shows the relaxation if
this is not allowed. The widths of these lines show the difference
between allowing self-interactions to be repressive or not.

\newpage

% Graph properties for different P threshold values.
\begin{center}
\rotatebox{0}{
\epsfig{figure=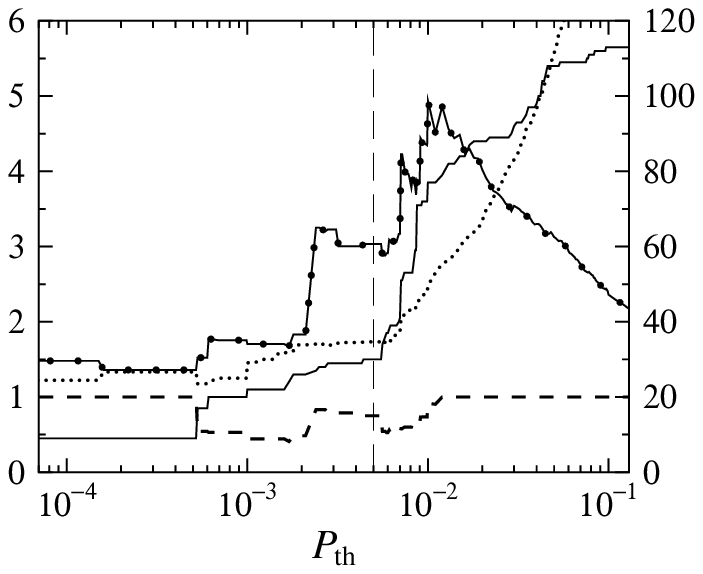,width=14cm}
}
\end{center}

\begin{center}
\vspace{10mm}
{\bf Fig.~1}
\end{center}

\newpage

% The Pth=0.005 network.
\begin{center}
\rotatebox{0}{
\epsfig{figure=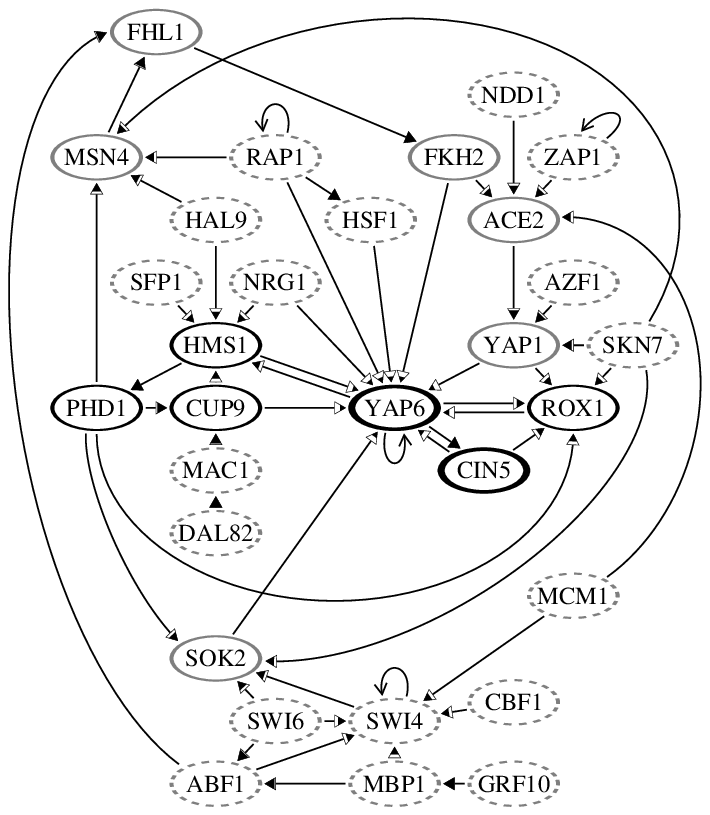,width=14cm}
}
\end{center}

\begin{center}
\vspace{10mm}
{\bf Fig.~2}
\end{center}

\newpage

% 3 Derrida plots (random,canalyzing}
\begin{center}
\rotatebox{0}{
\epsfig{figure=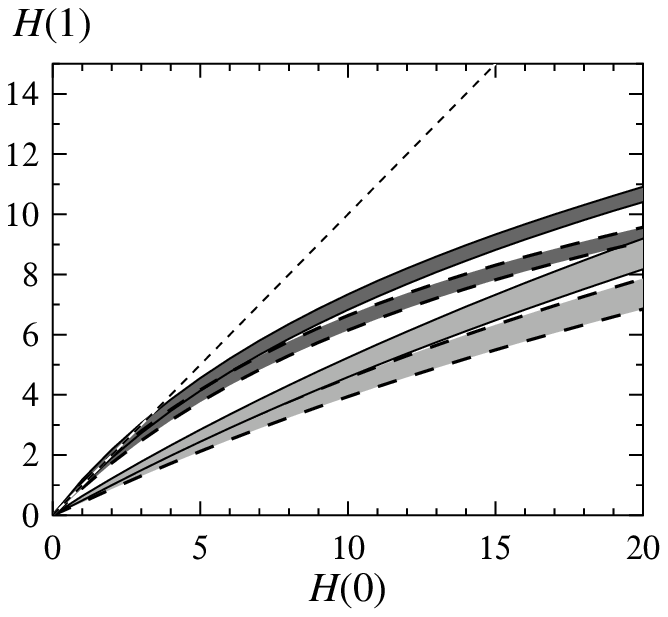,width=14cm}
}
\end{center}

\begin{center}
\vspace{10mm}
{\bf Fig.~3}
\end{center}

\newpage

% Time development of H from fixed point.
\begin{center}
\rotatebox{0}{
\epsfig{figure=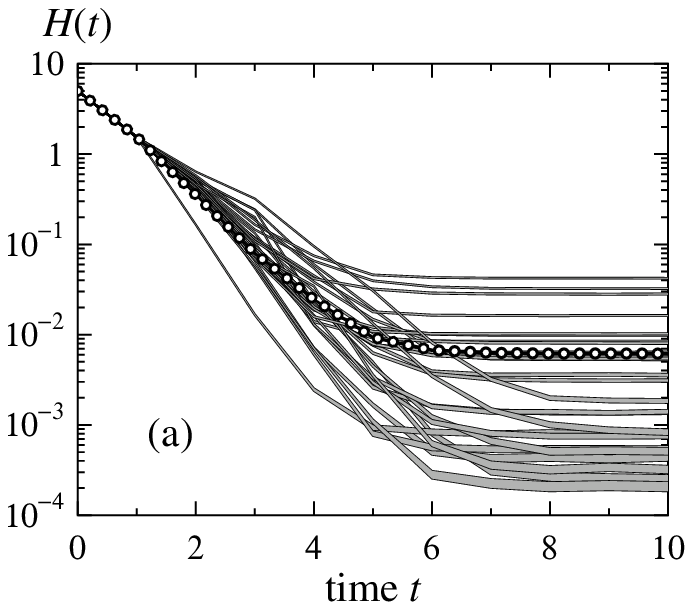,width=14cm}
}
\end{center}

\begin{center}
\vspace{10mm}
{\bf Fig.~4a}
\end{center}

%\newpage
%\mbox{}
\newpage

% Time development of H from fixed point.
\begin{center}
\rotatebox{0}{
\epsfig{figure=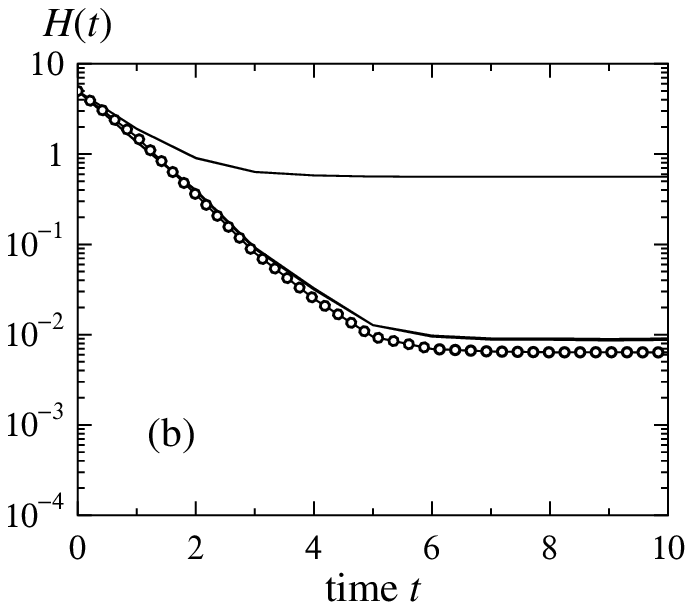,width=14cm}
}
\end{center}

\begin{center}
\vspace{10mm}
{\bf Fig.~4b}
\end{center}

\newpage

\subsection*{Supporting Information}

\hrule

%\vspace{0.05in}

{\it Random Boolean Network Models and the Yeast Transcriptional Network}\\
S. Kauffman,  C. Peterson, B. Samuelsson and C. Troein

%\vspace{0.05in}

{\bf Confronting Nested Canalyzing Functions with Compiled Data}

In order to compare compiled and generated distributions of rules, we
must ensure that every nested canalyzing function is always
represented by the same set of parameters $I_1, \ldots, I_K$ and $O_1,
\ldots, O_K$ (see Appendix in the printed article). All ambiguities in
the choice of the representation can be derived from the following
operations:
\begin{enumerate}
\item{The transformation $I_K \rightarrow \NOT I_K$ together with $O_K
\rightarrow \NOT O_K$ and $\Odef \rightarrow \NOT \Odef$.}
\item{Permutations among a set of inputs $i_m, \ldots, i_{m+p}$ such
that $O_m = \cdots = O_{m+p}$. The values of $I_m, \ldots, I_{m+p}$
are permutated in the same way as $i_m, \ldots, i_{m+p}$.}
\end{enumerate}

A unique representation is created from any choice of parameters in
two steps. First, 1.~is applied if $O_K \neq O_{K-1}$, which ensures
that $O_K = O_{K-1}$. In order to handle the special case $K=1$ in a
convenient way we define $O_0 = \FALSE$. Second, all intervals of
inputs $i_m,\ldots,i_{m+p}$ such that 2.~can be applied are identified
and permutated so that $I_m = \cdots = I_{m+q} = \FALSE$ and 
$I_{m+q+1} = \cdots = I_{m+p} = \TRUE$ for some $q$, $0 \leq q \leq p$.

\begin{figure}[tb]
\begin{center}
\epsfig{figure=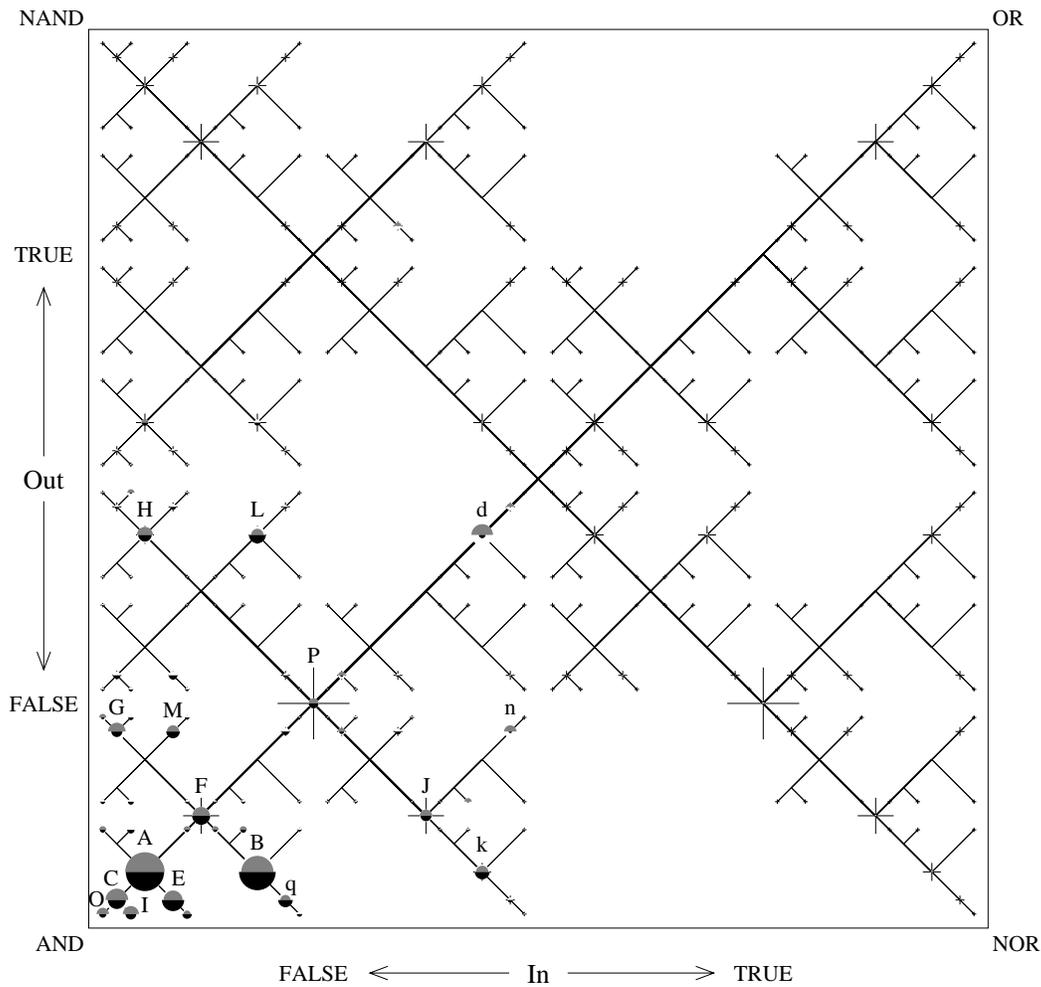,width=13.9cm}
\end{center}
\caption{
Compiled and generated rule distributions of nested canalyzing
functions. The gray half-circles have an area proportional to the
number of times each rule has been observed, while their black
counterparts reflect the calculated distribution. The labeled rules
are listed in Table 1. Capital labels mark rules that
are chain functions. Each rule is assigned a
coordinate in the unit square above (having $(0, 0)$ as its lower left
corner), according to $x = 1/2 + \sum_{m=1}^K 2^{-m}\phi(I_m)$, $y =
1/2 + \sum_{m=1}^K 2^{-m}\phi(O_m)$, where $\phi(\TRUE) = 1/2$ and
$\phi(\FALSE) = -1/2$. The crosses mark the possible coordinates for a
rule that is represented in its unique form. The lines indicate how
the coordinates can change when new inputs are added to an existing
rule.}
\label{fig: rules}
\end{figure}

Using the above described procedure, we can compare a generated rule
distribution with the compiled distribution. First, we
take away all redundant inputs of each observed rule. An input is
redundant if the output is never dependent on that input. 
Starting from 66, 45 and 22 nested canalyzing rules with 3, 4 and 5 inputs
respectively, the reduction renders
2, 9, 71, 35 and 16 such rules with 1, 2, 3, 4 and 5
inputs respectively. Second, we let $\alpha = 7$ and generate rule
distributions for each number of inputs. ($\alpha = 7$ is not based on
a precise fit, it was picked by hand to fit the distribution of $I_1,
\ldots, I_K$.) Table \ref{tab: rules} shows the result for the most
frequently observed rules, and Fig.~\ref{fig: rules} is a plot of
the full rule distribution. The calculated distribution fits
surprisingly well to the compiled one, considering that the model has
only one free parameter, $\alpha$.

\begin{table}[tbh]
\begin{center}
{\small
\begin{tabular}{crrll}
   & $\!\!\!\nobs\!\!$&$\!\!\!\ncalc\!\!$ & \canal{I_1}{O_1}, $\ldots$, \canal{I_K}{O_K}& Boolean expression \\
\hline
 A & 30 & 28 & \canal00, \canal00, \canal00                     & $i_1 \AND i_2 \AND i_3$\\
 B & 20 & 26 & \canal00, \canal00, \canal10	                & $i_1 \AND i_2 \AND \NOT i_3$\\
 C & 10 & 6  & \canal00, \canal00, \canal00, \canal00           & $i_1 \AND i_2 \AND i_3 \AND i_4$\\
 d &  9 & 1  & \canal00, \canal11, \canal11	                & $i_1 \AND (i_2 \OR i_3)$\\
 E &  7 & 10 & \canal00, \canal00, \canal00, \canal10           & $i_1 \AND i_2 \AND i_3 \AND \NOT i_4\!$\\
 F &  6 & 6  & \canal00, \canal00                               & $i_1 \AND i_2$\\
 G &  6 & 2  & \canal00, \canal00, \canal01, \canal01           & $i_1 \AND i_2 \AND \NOT (i_3 \AND i_4)\!$\\
 H &  5 & 4  & \canal00, \canal01, \canal01	                & $i_1 \AND \NOT (i_2 \AND i_3)$\\
 I &  5 & 2  & \canal00, \canal00, \canal00, \canal00, \canal10 & $i_1 \AND i_2 \AND i_3 \AND i_4$\\
   &    &    &                                                  & $~~~\AND \NOT i_5$\\
 J &  3 & 2  & \canal00, \canal10		                & $i_1 \AND \NOT i_2$\\
 k &  3 & 4  & \canal00, \canal10, \canal10	                & $i_1 \AND \NOT (i_2 \OR i_3)$\\
 L &  3 & 6  & \canal00, \canal01, \canal11	                & $i_1 \AND (\NOT i_2 \OR i_3)$\\
 M &  3 & 4  & \canal00, \canal00, \canal01, \canal11           & $i_1 \AND i_2 \AND (\NOT i_3 \OR i_4)$\\
 n &  3 & 0  & \canal00, \canal10, \canal11, \canal11           & $i_1 \AND \NOT i_2 \AND (i_3 \OR i_4)$\\
 O &  3 & 1  & \canal00, \canal00, \canal00, \canal00, \canal00 & $i_1 \AND i_2 \AND i_3 \AND i_4 \AND i_5$\\
 P &  2 & 2  & \canal00				                & $i_1$\\
 q &  2 & 4  & \canal00, \canal00, \canal10, \canal10           & $i_1 \AND i_2 \AND \NOT (i_3 \OR i_4)$
\end{tabular}
}
\end{center}
\caption{The list of nested canalyzing rules observed more than once
in [5]. $\nobs$ is the number of
observations in the compiled list of rules, whereas $\ncalc$ is
the average number of rules
in the generated distribution. Each rule is described both as an
ordinary Boolean expression, and with the
parameters $I_1, \ldots, I_K$ and $O_1, \ldots, O_K$, where $\Odef =
\NOT O_K$. 0 and 1 correspond to $\FALSE$ and $\TRUE$, respectively.
The labels serve as references in Fig.~1, and capital
labels mark rules that are chain functions. ($\NOT$has higher operator
precedence than {\sc and}, whereas the precedences of$\OR$and$\XOR$are
lower.)}
\label{tab: rules}
\end{table}

\begin{table}[tb]
\begin{center}
{\small
\begin{tabular}{rll}
$\!\!\!\nobs\!\!$ & \canal{I_1}{O_1}, $\ldots$, \canal{I_K}{O_K}& Boolean expression \\
\hline
 2 & \canal00, \canal00, \noncanal                    & $i_1 \AND i_2 \AND (\NOT i_3 \AND i_4$\\
   &                                                  & $~~~\OR \NOT i_4 \AND i_5)$\\
 1 & \canal01, \canal00, \canal00		      & $\NOT i_1 \OR i_2 \AND i_3$\\
 1 & \canal00, \canal10, \canal01, \canal01	      & $i_1 \AND \NOT (i_2 \OR i_3 \AND i_4)$\\
 1 & \canal00, \canal11, \canal00, \canal00	      & $i_1 \AND (i_2 \OR i_3 \AND i_4)$\\
 1 & \canal00, \canal11, \canal11, \canal11	      & $i_1 \AND (i_2 \OR i_3 \OR i_4)$\\
 1 & \canal01, \canal11, \canal00, \canal10	      & $\NOT i_1 \OR i_2 \OR i_3 \AND \NOT i_4$\\
 1 & \canal00, \canal00, \canal00, \canal01, \canal01 & $i_1 \AND i_2 \AND i_3 \AND \NOT (i_4 \AND i_5)$\\
 1 & \canal00, \canal00, \canal00, \canal10, \canal10 & $i_1 \AND i_2 \AND i_3 \AND \NOT (i_4 \OR i_5)$\\
 1 & \canal00, \canal00, \canal00, \canal11, \canal11 & $i_1 \AND i_2 \AND i_3 \AND (i_4 \OR i_5)$\\
 1 & \canal00, \canal00, \canal01, \canal01, \canal01 & $i_1 \AND i_2 \AND \NOT (i_3 \AND i_4 \AND i_5)$\\
 1 & \canal00, \canal00, \canal10, \canal01, \canal01 & $i_1 \AND i_2 \AND \NOT (i_3 \OR i_4 \AND i_5)$\\
 1 & \canal00, \canal00, \canal10, \canal11, \canal01 & $i_1 \AND i_2 \AND \NOT i_3 \AND (i_4 \OR \NOT i_5)\!$\\
 1 & \canal00, \canal01, \canal01, \canal01, \canal11 & $i_1 \AND \NOT (i_2 \AND i_3 \AND i_4$\\
   &                                                  & $~~~\AND \NOT i_5)\!$\\
 1 & \canal00, \canal10, \canal11, \canal00, \canal10 & $i_1 \AND \NOT i_2 \AND (i_3 \OR i_4 \AND \NOT i_5)\!$\\
 1 & \canal00, \canal00, \noncanal                    & $i_1 \AND i_2 \AND (i_3 \XOR i_4)$\\
 1 & \canal00, \noncanal                              & $i_1 \AND (i_2 \XOR i_3 \AND i_4)$\\
 1 & \canal00, \noncanal                              & $i_1 \AND (2\leq)(i_2, i_3, \NOT i_4)$\\
 1 & \canal10, \noncanal                              & $\NOT i_1 \AND (i_2 \AND \NOT i_3$\\
   &                                                  & $~~~\OR i_3 \AND \NOT(i_4 \OR i_5))$\\
\end{tabular}
}
\end{center}
\caption{Continuation of Table 1, containing the remainder of
rules listed in [5]. The Boolean function $(2\leq)$ is $\TRUE$
if at least two of its arguments are $\TRUE$.} 
%$\nobs$ is the number of occurrences, and the rules are
%described both as an ordinary Boolean expression, and with the
%parameters $I_1, \ldots, I_K$ and $O_1, \ldots, O_K$,
%where $\Odef = \NOT O_K$. 0 and 1 correspond to $\FALSE$ and $\TRUE$,
%respectively.
%The Boolean function $(2\leq)$ is $\TRUE$ if at least two of its
%arguments are $\TRUE$. ($\NOT$has higher operator precedence
%than {\sc and}, whereas the precedences of$\OR$and$\XOR$are lower.)}
\label{tab: rest of rules}
\end{table}

\end{document}